\documentclass[pra,twocolumn]{revtex4}
\usepackage{amsmath,amssymb,mathrsfs}
\usepackage{psfrag}
\usepackage{graphicx}
\usepackage{graphics}
\usepackage{epsfig}
\usepackage{bm}
\usepackage{color}
\usepackage{verbatim,color,ulem}

\newcommand{\beq}{\begin{equation}}
\newcommand{\eeq}{\end{equation}}
\newcommand{\beqa}{\begin{eqnarray}}
\newcommand{\eeqa}{\end{eqnarray}}
\newcommand{\ba}{\begin{aligned}[b]}
\newcommand{\ea}{\end{aligned}}

\begin{document}

\title{Bose-Einstein Condensation, Fluctuations and Spontaneous Symmetry Breaking}

\begin{abstract}
The realisation of Bose-Einstein condensation under grand-canonical
conditions has provided the experimental evidence for the simultaneous
occurrence of macroscopic fluctuations and phase coherence of the
condensate. The observation of these two features, against a
consolidated tradition which wants the fluctuations to be pathological
(grand canonical catastrophe) and incompatible with spontaneous
symmetry breaking, calls for a comprehensive rethinking
of the approach to the problem. In this paper we consider the uniform
ideal gas in a box and we present an alternative conceptual framework.
We show that the usually-employed Bogoliubov quasi-average
construction fails to reproduce the broken-symmetry state. The
observed features are accounted for by a different pattern of
spontaneous symmetry breaking, characterised by condensation of
fluctuations and long-range correlations of the order parameter.
\end{abstract}

\author{A. Crisanti}
\email{andrea.crisanti@uniroma1.it}
\affiliation{Dipartimento di Fisica, Universit\`a di Roma Sapienza, P.le Aldo Moro 2, 00185 Rome, Italy}
\affiliation{Istituto dei Sistemi Complessi - CNR, P.le Aldo Moro 2, 00185 Rome, Italy}

\author{A. Sarracino}
\email{alessandro.sarracino@unicampania.it}
\affiliation{Dipartimento di Ingegneria, Universit\`a della Campania ``Luigi Vanvitelli'', via Roma 29, 81031 Aversa (CE), Italy}

\author{M. Zannetti}
\email{mrc.zannetti@gmail.com}
\affiliation{Dipartimento di Fisica ``E. R. Caianiello'',
Universit\`a di Salerno, via Giovanni Paolo II 132, 84084 Fisciano (SA), Italy,}

\maketitle

\section{INTRODUCTION}

A century after its discovery, and even in the simplest setting of the ideal gas in the grand-canonical ensemble (GCE), 
Bose-Einstein condensation (BEC) remains not fully understood.
The derivation is exact and by now standard textbook material. Yet, we are warned that what we learn
should not be entirely trusted. The reason is that condensation is accompanied
by macroscopic fluctuations, an outcome disturbing to the point of being tellingly dubbed as
the {\it grand-canonical catastrophe} (${\cal GCC}$) \cite{Ziff,Holthaus,Fujiwara,Grossmann,Weiss}. Some of the motivations for the refusal to accept a seemingly straightforward
result are difficult to address, being grounded on vague notions
like the extravagance of such large fluctuations.
For a rich account of this sort of criticism
see the recent review by Kruk et al. and references therein \cite{Kruk}. 

On a different level is the objection based on the alleged incompatibility of the condensate fluctuations 
with the spontaneous breaking of the $U(1)$ symmetry (SSB)~\cite{Yukalov} and this needs to be addressed. The  connection between BEC and SSB has
been established in the strong sense of an equivalence~\cite{Suto,Lieb}.
However, the chain of reasoning rests on the key assumption that the broken-symmetry state can be faithfully
reproduced via the Bogoliubov
quasi-average construction~\cite{Bogoliubov2}. The problem with this is that the quasi-average scheme 
implies the exactness of the Bogoliubov $c$-number substitution which, in turn, excludes that there might be
fluctuations of the condensate. Hence, 
according to the conventional theoretical picture, based on quasi averages,
there is an incompatibility between SSB and ${\cal GCC}$ which is resolved by regarding the 
latter as a pathological feature of the GCE.
Remarkably, this attitude has persisted even after
the landmark experimental realisation of BEC in a photon system has demonstrated 
both that the ${\cal GCC}$ is a {\it real} phenomenon observable in the lab~\cite{Klaers,Schmitt1} {\it and}  
that it is compatible with SSB~\cite{Schmitt2}.

It is then apparent that the usual approach to the BEC-SSB connection needs some overhauling. 
So, we reconsider the whole problem from scratch. Taking as stepping stone the fact that
the validity of the ${\cal GCC}$ does not depend on symmetry considerations (see below) and building on ideas proposed in previous works~\cite{MZ,CSZ}, in this paper we show 
that the apparent incompatibility with SSB descends right from how SSB has been conventionally treated.
We trace back the source of the problem to the uncritical use of the quasi-average
procedure. Specifically, we show that, when applied in the GCE,
the external-field perturbation involved in the quasi average is singular in the condensed phase 
and produces a broken-symmetry state
which is {\it not equivalent} to the unperturbed one. Then, we propose to extract the latter, 
which is the physically-meaningful one, from the set of coherent states entering 
the Glauber-Sudarshan P-representation of the density matrix \cite{Glauber,Sudarshan}. In addition to account for the 
experimentally-established compatibility of ${\cal GCC}$ and SSB, 
the condensed phase is now found to be characterised
by the condensation of fluctuations of the matter field,
implying the existence of long range correlations. 
This is a radical departure from the conventional picture, which predicts simple
ordering of the matter field, like in a ferromagnet, with negligible fluctuations and correlations.

We stress that the constraint enforced 
by fixing the average density (a key step in the derivation of BEC in the GCE) amounts to 
a mean-field approximation, whose instability
to external-field perturbations is well known from other instances of the same 
approximation \cite{Lewis1,Lewis2,KT,YW,Castellano,Fusco}.
In this respect, the photons
experiment is all the more important because provides the so-far-unique
experimental platform realising a theoretical scheme which in other contexts, such as magnetic systems, is
elusive if not impossible to observe.

\section{BEC AND GRAND CANONICAL CATASTROPHE}

We are concerned with a system of non-interacting  bosons of mass $m$ described by
the second-quantized Hamiltonian 
\beq 
{\hat {\cal H}} = \int d{\bf r} \ {\hat \psi}^{\dagger}({\bf r}) 
\left[ -\frac{\hbar^2}{2m} \nabla^2 + U({\bf r}) 
\right] {\hat \psi}({\bf r}).
\label{hamilton.0}
\eeq 
We shall first work out the problem in the simpler case of the gas in a
$3d$ box of volume $V=L^3$, setting the confining potential $U({\bf r}) \equiv 0$. The experimentally relevant case of the gas 
in the harmonic trap will be briefly discussed at the end.
Imposing periodic boundary conditions, the energy eigenfunctions are the plane waves
$\phi_{\bf k}({\bf r}) = \frac{1}{\sqrt{V}}\, e^{i{\bf k} \cdot {\bf r}}$ with wave vectors $k_{x,y,z} = \frac{2\pi}{L}(0,\pm 1,..)$
and eigenvalues $\epsilon_{\bf k}= (\hbar^2 k)^2/2 m$. Expanding the field operator 
${\hat \psi}({\bf r}) = \sum_{{\bf k}}  {\hat a}_{\bf k} \phi_{\bf k}({\bf r})$, 
where ${\hat a}_{\bf k}$ and ${\hat a}_{\bf k}^{\dagger}$ are the annihilation and creation operators in the one-particle
state $| {\bf k} \rangle$, the Hamiltonian goes into the diagonal form
${\hat {\cal H}} = \sum_{{\bf k}} {\hat {\cal H}}_{{\bf k}}$, with
${\hat {\cal H}}_{{\bf k}} = \epsilon_{\bf k} {\hat a}_{\bf k}^{\dagger}{\hat a}_{\bf k}$.
The Gibbs density matrix factorises $\hat{{\cal D}} = \prod_{{\bf k}} \hat{{\cal D}}_{{\bf k}}$, with
\beq
\hat{{\cal D}}_{\bf k} = 
\frac{1}{\cal Z}_{\bf k} \exp \{-(\beta \hat{\cal H}_{\bf k} + \alpha {\hat n}_{\bf k})\}, \quad \alpha=-\beta \mu,
\label{dens.01}
\eeq
where $\beta$ is the inverse temperature, $\mu$ is the chemical potential,
$\hat{n}_{\bf k} =  \hat{a}_{\bf k}^\dagger \hat{a}_{\bf k}$ and
${\cal Z}_{\bf k}$ is obtained from ${\rm Tr} \,  \hat{{\cal D}}_{\bf k} = 1$.

Denoting averages by angular brackets, 
a straightforward calculation of the ${\bf k}$-mode occupation density 
$\rho_{\bf k} = \frac{1}{V} \langle {\hat n}_{\bf k} \rangle$
and of the mean-square fluctuations $\delta^2\rho_{\bf k} = \frac{1}{V^2} [\langle {\hat n}^2_{\bf k}\rangle - \langle {\hat n}_{\bf k} \rangle^2]$, 
yields
\beq
\rho_{\bf k} = \frac{1}{V[e^{\beta \epsilon_{\bf k} + \alpha} -1]},  \quad
\delta^2\rho_{\bf k} =  \rho^2_{\bf k} +  \frac{1}{V}\rho_{\bf k}.
\label{dens.10}
\eeq
Upon varying the control parameters $(\beta,\alpha,V)$ the behaviour of these quantities is smooth and featureless,
as it should be in a free theory.
So, in order to get not-trivial results some kind of interaction must be introduced. 
In the GCE this is covertly done by solving  the equation
for the total density $\rho = \sum_{\bf k} \rho_{\bf k}(\alpha)$ with respect to $\alpha$
and by substituting the solution $\alpha(\beta,\rho,V)$ wherever $\alpha$ appears
(see for example Refs.~\cite{Ziff,Huang}).
Trading $\alpha$ with $\rho$ as a control parameter is the crucial step for the appearance of the BEC singularity. 
If the actual density (not the averaged one) had
been kept fixed (canonical ensemble), an obvious coupling would have been introduced
through the sum rule $\rho = \sum_{\bf k} \rho_{\bf k}$. 
In the GCE, instead, by fixing the average density
the constraint is softened leaving the structure formally
non-interacting.
Nevertheless, a coupling is still implemented via the self-consistency 
relation $\rho = \sum_{\bf k} \rho_{\bf k}(\rho)$, which is characteristic
of the mean-field approximations. Particularly relevant in the present context is the connection with the mean spherical model~\cite{Lewis1,Lewis2,KT,CSZ,YW,Lewis}.

Separating the ground-state contribution $\rho_{\bf 0}$ from
the excited states $\rho^\prime = \sum_{{\bf k} \neq 0} \rho_{\bf k}$ and replacing this sum
with an integral, the equation for $\alpha$ reads
\beq
\rho=\rho_{\bf 0}(\alpha) + \lambda^{-3}g_{3/2}(\alpha),
\label{BEC.001}
\eeq
where $g_{3/2}(\alpha)$ is the Bose function~\cite{Ziff} and $\lambda = \sqrt{2\pi \beta \hbar^2/m}$ is the de Broglie thermal wave length. Solving, substituting the solution $\alpha(\beta,\rho,V)$ into Eq.~(\ref{dens.10}) 
and taking the $V \to \infty$ limit, one finds BEC and, inevitably, the concomitant ${\cal GCC}$
\beq
\rho_{\bf 0}(\rho) = \left \{ \begin{array}{ll}
0, \quad $for$ \quad \Delta \rho \leq 0, \\
\\
\Delta \rho, \quad $for$ \quad \Delta \rho > 0,  
        \end{array}
        \right . 
        \quad 
\label{ro.1}
\eeq

\beq
\delta^2\rho_{\bf 0}(\rho)= \left \{ \begin{array}{ll}
0, \quad $for$ \quad \Delta \rho \leq 0, \\
\\
\Delta \rho^{2}, \quad $for$ \quad \Delta \rho > 0,
        \end{array}
        \right . 
 \label{BEC.002}
  \eeq
where $\Delta \rho = \rho - \rho_c$, $\rho_c = \lambda^{-3} \, \zeta(3/2)$ is the critical density, 
and $\zeta$ is the 
Riemann zeta function. 
In the standard treatment of BEC the
transition is driven by $\beta$, keeping $\rho$ fixed. Here, by analogy with the photons' experiment,
$\rho$ is used as the driving parameter keeping $\beta$ fixed. The normal phase ($\Delta \rho < 0$) and
the condensed phase ($\Delta \rho > 0$) are separated by a critical point (see below) at $\Delta \rho = 0$.

\section{SPONTANEOUS SYMMETRY BREAKING}

Given the simplicity of the algebra involved, it is
hard to see why the result for $\delta^2 \rho_{\bf 0}$ ought to be discarded as pathological. 
Nonetheless, as previously recounted, it has been challenged invoking SSB~\cite{Yukalov}. Restricting, for simplicity, 
 the attention to the ${\bf 0}$-mode, the statement is that in the condensed
phase and in the thermodynamic limit the invariance of the density matrix $\hat{{\cal D}}_{\bf 0}$ under the gauge transformation
${\hat a}_{\bf 0} \mapsto e^{i\theta} \, {\hat a}_{\bf 0}$ is spontaneously broken. Then, 
$\hat{{\cal D}}_{\bf 0}$ must be regarded as the mixture of dynamically-disjoint ergodic components~\cite{Roepstorff,WZ}
\beq
\hat{\cal D}_{\bf 0} = \frac{1}{2\pi} \int_0^{2\pi} d\theta \, \hat{\cal D}_{{\bf 0},\theta} \, ,
\label{decomp.1}
\eeq
each of which with a definite phase $\theta$. Only one of these represents the physically-meaningful state,
since the real system in the lab remains confined in one component~\cite{Schmitt2}. Consequently, the meaningful averages
are those taken with $\hat{\cal D}_{{\bf 0},\theta}$ whose form, at this stage, is not known and needs to be determined. 

Here, there are three distinct issues: first, establish if SSB takes place, then, if so, 
find out what the form of $\hat{\cal D}_{{\bf 0},\theta}$
is and, finally, check if on averaging with $\hat{\cal D}_{{\bf 0},\theta}$ the ${\cal GCC}$ nuisance is disposed of.
Actually, the last one is the easiest to deal with, since symmetrical operators, like $\hat{n}_{\bf 0}$ and $\hat{n}^2_{\bf 0}$,
are insensitive to the phase. Therefore, it doesn't matter whether the average is taken with the 
symmetric matrix $\hat{\cal D}_{\bf 0}$ or with
one of the components $\hat{\cal D}_{{\bf 0},\theta}$. The outcome is the same. Consequently, the results~(\ref{ro.1},\ref{BEC.002})
for $\rho_{\bf 0}$ and $\delta^2 \rho_{\bf 0}$ hold also in the broken-symmetry state, whatever this might be,
implying that the
${\cal GCC}$ has nothing to do with SSB. Should it disappear, as indeed it does with 
the Bogoliubov procedure, it can only mean that the quasi-average construction fails to reproduce  
$\hat{\cal D}_{{\bf 0},\theta}$.

Let us now see how this method works, since it is widely used to address the first and second issue. The $U(1)$ symmetry is
explicitly broken by applying
a complex  external field $\nu = |\nu|e^{i\theta_\nu}$, which 
couples to the order parameter ${\hat \psi_{\bf 0}} = \hat{a}_{\bf 0} \phi_{\bf 0}({\bf r}) = \frac{1}{\sqrt{V}} \hat{a}_{\bf 0}$.
There are no changes for ${\bf k} \neq {\bf 0}$, while for ${\bf k} = {\bf 0}$ the Hamiltonian and density matrix go into
$\hat{{\cal H}}^{(\nu)}_{{\bf 0}} = \hat{{\cal H}}_{{\bf 0}}  -
V (\nu {\hat \psi_{\bf 0}}^{\dagger} + \nu^* {\hat \psi_{\bf 0}})$
and
$\hat{\cal D}_{\bf 0}^{(\nu)} = \frac{1}{\cal Z}_{\bf 0} \, e^{-\beta \hat{{\cal H}}_{\bf 0}^{(\nu)} - \alpha \hat{n}_{\bf 0}}$.  
Clearly, if $\nu \to 0$ the unbiased state $\hat{\cal D}_{\bf 0}$ is recovered. However, the idea is that if the removal of $\nu$
is done {\it after} taking the $V \to \infty$ limit {\it and} if SSB has occurred, then the $\nu \to 0$ limit of 
$\hat{\cal D}_{\bf 0}^{(\nu)}$ remains locked into the particular component $\hat{\cal D}_{{\bf 0},\theta_\nu}$
selected by the phase of the biasing field. The occurrence of SSB is then revealed by a non-null value of the quasi-averaged 
order parameter
$\langle {\hat \psi_{\bf 0}} \rangle^{({\rm qa})}_{\theta_\nu} =\lim_{\nu \to 0} \lim_{V \to \infty} \langle {\hat \psi_{\bf 0}} \rangle^{(\nu)}$.
The order of limits is crucial. The reverse order produces the regular average~\cite{Bogoliubov2}
$\langle {\hat \psi_{\bf 0}}\rangle^{({\rm ra})} = \lim_{V \to \infty} \,  \lim_{\nu \to 0} \langle {\hat \psi_{\bf 0}}\rangle^{(\nu)}$,
which vanishes by symmetry. In addition to checking on SSB,  the quasi-average method 
is usually also tacitly extended to the more delicate
task of constructing the 
ergodic-component $\hat{\cal D}_{{\bf 0},\theta}$ entering the decomposition~(\ref{decomp.1}),
by assuming that one can make the identification
\beq
\langle \cdot \rangle_{\theta}^{({\rm ra})} \stackrel{?}{=} \langle \cdot \rangle^{({\rm qa})}_{\theta},
\label{iden.1}
\eeq
where the left-hand side stands for the regular average taken in the (so far unknown) ergodic
component with the phase $\theta$. The catch, here, is that the above identification holds only if the 
external field's perturbation is not singular, which, as we shall see, is not the case with the
ideal gas in the GCE.

The issue is settled by resorting to the Glauber-Sudarshan 
P-representation~\cite{Glauber,Sudarshan} of the biased density matrix
$\hat{\cal D}^{(\nu)}_{\bf 0} = \int d^2 z \, P^{(\nu)} (z) \, |z \rangle \langle z |$,
where the coherent state $|z \rangle$ is the eigenvector of the volume-rescaled annihilation operator
${\hat \psi}_{\bf 0}$, with the complex eigenvalue
$z = |z|e^{i\theta}$.
Averages of normally-ordered products are given by~\cite{Glauber}
\beq
\langle {\hat \psi}_{\bf 0}^{\dagger n}{\hat \psi}^m_{\bf 0} \rangle^{(\nu)} =
\int d^2 z \, P^{(\nu)}(z) \, z^{*n} z^m.
\label{avgs.1}
\eeq 
The weight function reads~\cite{Mehta}
\beq
P^{(\nu)}(z) =\frac{1}{\pi \sigma} \,
\exp \left \{ -\frac{1}{\sigma} (z^* - \gamma^*) (z - \gamma)\right \},
\label{GS.2}
\eeq
where
\beq
\sigma = [V \left (e^\alpha- 1 \right )]^{-1}, \quad 
\gamma = \beta \nu/ \alpha,
\label{GS.3}
\eeq
from which follows
\beq
\langle \hat{\psi}_{\bf 0} \rangle^{(\nu)}= \gamma,  \quad \rho^{(\nu)}_{\bf 0} =  |\langle \hat{\psi}_{\bf 0} \rangle^{(\nu)}|^2 + \sigma.
\label{GS.4}
\eeq
Again, it is necessary to first solve for $\alpha$ as a function of $(\beta,\rho,V,\nu)$. 
This is done in the Appendix obtaining a result which depends on the order of limits, thus
revealing the singular nature of the perturbation.  
Inserting the outcome into Eq.~(\ref{GS.2}) and using Eq.~(\ref{GS.3}), in the quasi-average case we find
\beq
\lim_{\nu \to 0} \lim_{V \to \infty}  P^{(\nu)}(z)= P^{({\rm qa})}_{\theta_\nu}(z)  = \left \{ \begin{array}{ll}
\delta^2(z), \quad $for$ \quad \Delta \rho \leq 0, \\
\\
\delta(\theta - \theta_\nu) \, \delta(|z| - \sqrt{\Delta \rho}), \\\quad $for$ \quad \Delta \rho > 0.
        \end{array}
        \right . 
 \label{P0.1}
  \eeq
Hence, in the condensed phase 
the density matrix reduces to the projector on the single coherent state selected by the phase of the vanishing external field~\cite{CR}
\beq
\hat{\cal D}_{{\bf 0},\theta_\nu}^{({\rm qa})} = | e^{i\theta_\nu}\sqrt{\Delta \rho}\rangle \langle e^{i\theta_\nu}\sqrt{\Delta \rho}|.
\label{ra.102}
\eeq
Conversely, the regular-average procedure yields 
\beq
\lim_{V \to \infty} \lim_{\nu \to 0}  P^{(\nu)}(z) = P^{({\rm ra})}(z)  = \left \{ \begin{array}{ll}
\delta^2(z), \quad $for$ \quad \Delta \rho \leq 0, \\
\\
\frac{1}{\pi} \, \frac{1}{\Delta \rho} \exp \left \{ - \frac{|z|^2}{\Delta \rho} \right \}, \\\quad $for$ \quad \Delta \rho > 0, 
        \end{array}
        \right . 
 \label{P0.2}
  \eeq
i.e. a distribution uniform over the phases and spread-out over the amplitudes.
Therefore, the unperturbed broken-symmetry state entering the decomposition~(\ref{decomp.1}),
which we reiterate is the physically meaningful one,  
can be readily obtained simply by omitting the phase integration and obtaining
\beq
\hat{\cal D}_{\bf 0,\theta}^{({\rm ra})} = 2\int_0^\infty d |z| \, |z| \, 
\frac{1}{\Delta \rho} \exp \left \{ - \frac{|z|^2}{\Delta \rho} \right \}
| e^{i\theta}|z| \rangle \langle e^{i\theta}|z||.
\label{ra.105}
\eeq
Comparing Eqs.~(\ref{ra.102}) and~(\ref{ra.105}), it is evident 
that the physical broken-symmetry state cannot be constructed by means of the quasi-average
and that Eq.~(\ref{iden.1}) does not hold.
Carrying out the calculation of Eq.~(\ref{avgs.1}), in the two cases we have 
\beqa
& & \langle {\hat \psi}_{\bf 0}^{\dagger n} {\hat \psi}_{\bf 0}^m \rangle^{({\rm qa})}_ \theta = 
e^{i(m-n)\theta} \,\Delta \rho^{(n+m)/2}, 
\label{gen.1} \\
& &\langle {\hat \psi}_{\bf 0}^{\dagger n} {\hat \psi}_{\bf 0}^m \rangle^{({\rm ra})}_ \theta = 
\Gamma((n+m+2)/2) \, e^{i(m-n)\theta} \,\Delta \rho^{(n+m)/2}, \nonumber \\
\label{gen.2}
\eeqa
where $\Gamma$ is the Euler gamma function. In particular, for $(n=0,m=1)$ we get 
\beqa
& & \langle {\hat \psi}_{\bf 0} \rangle^{({\rm qa})}_ {\theta_\nu} = e^{i\theta_\nu} \,\sqrt{\Delta \rho},
\label{opp.1} \\
& & \langle {\hat \psi}_{\bf 0} \rangle^{({\rm ra})}_{\theta}  = 
\Gamma(3/2) e^{i\theta} \sqrt{\Delta \rho}, 
\label{opp.2}
\eeqa
with $\Gamma(3/2)= \sqrt{\pi}/2 < 1$. Hence,
the $c$-number substitution 
${\hat \psi}_{\bf 0} \mapsto c=\langle {\hat \psi}_{\bf 0} \rangle^{({\rm qa})}_ {\theta_\nu}$ is exact in the quasi-average case
$\langle {\hat \psi}_{\bf 0}^{\dagger n} {\hat \psi}_{\bf 0}^m \rangle^{({\rm qa})}_ {\theta_\nu} =
\langle {\hat \psi}_{\bf 0}^{\dagger} \rangle^{({\rm qa})n}_ {\theta_\nu} \langle {\hat \psi}_{\bf 0} \rangle^{({\rm qa})m}_ {\theta_\nu}$,
but does not hold in the regular-average case because of the gamma functions.
It is important to note that the $\delta$-functions in Eq.~(\ref{P0.1}) match the
P-representation of the density matrix in the canonical ensemble~\cite{CSZ} and, therefore, that the
quasi-average ensemble is equivalent to the canonical one implying, in turn, that the 
lack of equivalence between the quasi-average and the regular-average is 
the same as the one between the canonical and grand canonical
ensembles. Moreover, the instability under the external perturbation
 is a feature which further underscores
the mean-field character of the formal apparatus, since it is well known to occur in the already 
mentioned mean-spherical model~\cite{KT,YW} and in the large-$N$ limit~\cite{Castellano,Fusco}, 
where $V \to \infty$ and $h \to 0$ do not commute, $h$ being the magnetic field.

\section{CONDENSATION OF FLUCTUATIONS}

A major consequence of the discrepancy between regular and quasi averages concerns the nature of the condensed phase,
specifically the mechanism of formation of the condensate at the underlying level of the field ${\hat \psi}_{\bf 0}$.
Setting $(n=1,m=1)$, in both cases we recover the same result~(\ref{ro.1})
$\rho_{\bf 0} = \langle {\hat \psi}^\dagger_{\bf 0} {\hat \psi}_{\bf 0} \rangle = \Delta \rho$.
Since in the quasi-average case this is achieved by developing the appropriate
anomalous average
$\rho_{\bf 0}^{({\rm qa})} = 
\langle {\hat \psi}^\dagger_{\bf 0} \rangle^{({\rm qa})}_\theta \langle {\hat \psi}_{\bf 0} \rangle^{({\rm qa})}_\theta$,
BEC is an instance of {\it ordering transition}, like the ferromagnetic transition.
Conversely, from Eq.~(\ref{opp.2}) follows
\beq
\rho_{\bf 0}^{({\rm ra})} =
\Delta \rho = 
\langle {\hat \psi}^\dagger_{\bf 0} \rangle^{({\rm ra})}_\theta \langle {\hat \psi}_{\bf 0} \rangle^{({\rm ra})}_\theta
 + \Phi^{({\rm ra})},
        \label{av.201}
        \eeq
where, next to the ordering piece, there appears also the mean-square fluctuation of ${\hat \psi}_{\bf 0}$
\beq        
\Phi^{({\rm ra})} = \langle \delta {\hat \psi}^\dag_{\bf 0}\delta {\hat \psi}_{\bf 0} \rangle^{({\rm ra})}_\theta = [1-\pi/4]{\Delta \rho}, 
\quad \delta {\hat \psi}_{\bf 0} = {\hat \psi}_{\bf 0} - \langle {\hat \psi}_{\bf 0} \rangle^{({\rm ra})}_\theta.
\label{av.201bis}
        \eeq
Therefore, ordering is not enough to build up the condensate. The missing piece must be supplemented by 
the fluctuations of the amplitude mode.  
Hence, in this case BEC is a different kind of transition involving
{\it condensation of fluctuations}, in the sense that fluctuations of macroscopic size are not the cumulative
effect contributed by many degrees of freedom, but are produced by 
a {\it single} degree of freedom~\cite{MZ}. This is a phenomenon
of general occurrence, observed in a wide range of systems~\cite{Castellano,Corberi,Corberi2,Corberi3,Filiasi,Ferretti}.
Then, it must be stressed that BEC in the GCE
involves condensation in two different ways, which must be kept distinct: next to the usual condensation
in the sense of a macroscopic number of bosons accumulating in the ground state, there is
also condensation of fluctuations at the underlying microscopic
level of the order parameter ${\hat \psi}_{\bf 0}$, which is indispensable
to meet the required size of the condensate $\rho_{\bf 0}$ and whose mesoscopic manifestation is in the ${\cal GCC}$.

However, this is not yet the root cause of ${\cal GCC}$. Abiding by the rule of thumb
that wants large fluctuations due to long-range correlations, we must search for these
correlations.
Separating the ground-state from the excited-states contribution
${\hat \psi}({\bf r}) = {\hat \psi}_{\bf 0}({\bf r}) + {\hat \psi}^\prime({\bf r})$,
the first-order {\it connected} correlation function
$G_{\rm c}({\bf r} - {\bf r}^\prime) = \langle \hat{\psi}^\dagger({\bf r}) \hat{\psi}({\bf r}^\prime) \rangle
-\langle \hat{\psi}^\dagger({\bf r}) \rangle \langle \hat{\psi}({\bf r}^\prime) \rangle$
splits into the sum
$G_{\rm c}({\bf r} - {\bf r}^\prime) = G_{{\rm c},{\bf 0}}({\bf r} - {\bf r}^\prime) + G_{\rm c}^\prime({\bf r} - {\bf r}^\prime)$.
The first reads $G_{{\rm c},{\bf 0}}({\bf r} - {\bf r}^\prime) =
\langle \delta\hat{\psi}_{\bf 0}^\dagger  \delta\hat{\psi}_{\bf 0} \rangle$, whose form
in the condensed phase depends on how the average is taken
\beq
G_{{\rm c},{\bf 0}}({\bf r} - {\bf r}^\prime)= \left \{ \begin{array}{ll}
0, \quad $quasi average$, \\
\Phi^{({\rm ra})}, \quad $regular average$.
        \end{array}
        \right .
        \label{gzero.1}
        \eeq
The second does not depend on the way of averaging and for
$|{\bf r} - {\bf r}^\prime|$ and $V$ large enough is given by~\cite{Gunton,Ziff}
$G_{\rm c}^\prime({\bf r} - {\bf r}^\prime) =  \frac{\exp \{-|{\bf r} - {\bf r}^\prime|/\xi\}}{\lambda^2 |{\bf r} - {\bf r}^\prime|}$,
with $\xi =\frac{1}{\lambda^2|\Delta \rho|}$ for $\Delta \rho < 0$ and $\xi = \infty$ for $\Delta \rho \geq 0$.
Hence, in the normal phase there is one fluctuating field ${\hat \psi}({\bf r})$
with a finite correlation length, which diverges upon approaching the critical point at $\Delta \rho =0$~\cite{Gunton,Reyes}. 
In the condensed phase both fields
$\delta {\hat \psi}_{\bf 0}({\bf r})$ and ${\hat \psi}^\prime({\bf r})$ are critical but belong to different
universality classes, as revealed by the power-law decays 
\beq
G_{{\rm c},{\bf 0}}({\bf r} - {\bf r}^\prime) =  \frac{\Phi^{({\rm ra})}}{|{\bf r} - {\bf r}|^{a_0}}, \quad
G_{\rm c}^\prime({\bf r} - {\bf r}^\prime)  =  \frac{1}{\lambda^2 |{\bf r} - {\bf r}^\prime|^{a^\prime}},
\label{conn.3}
\eeq
with the different exponents $a_0=0$ and $a^\prime=1$. Correlations non-decaying with the
distance seems to be a characteristic feature of condensation of fluctuations. It has been investigated
in detail in the context of the Ising model with anti-periodic boundary
conditions, which exhibits condensation of fluctuations below the critical temperature,
in place of the usual ferromagnetic transition~\cite{FCZ1,FCZ2}. In that case, the spin-spin correlation
function remains constant and
the vanishing of the exponent $a$ has a nice geometrical interpretation
in terms of the Coniglio-Klein correlated clusters~\cite{CK}.

\section{CONCLUDING REMARKS}

As mentioned in the introductory remarks, 
the broad features of the scenario presented above
are not just a mere theoretical possibility, but have been experimentally observed in a photonic quantum gas. 
By realising BEC in GCE conditions in a dye-filled cavity~\cite{Klaers,Schmitt1}, evidence
has been produced for the existence of macroscopic fluctuations
of the condensate, consistently with the ${\cal GCC}$. 
Remarkably, in the same system, phase coherence of the condensate has been detected
on a time scale which grows linearly with the size of the condensate~\cite{Schmitt2}, indicating the occurrence 
of SSB in the thermodynamic limit and its compatibility with ${\cal GCC}$.
The ideal limit of this experimental system is represented
by a perfect gas of photons, endowed with an effective non-null mass and confined
in a trap modeled by the two-dimensional harmonic potential
$U({\bf r})=\frac{1}{2} m_{\textrm{eff}}\Omega^2 |{\bf r}|^2$.
Carrying out the calculations in the GCE, both
BEC and ${\cal GCC}$ are found(see for example Ref.~\cite{CSSZ}).
However, there are important formal differences with the gas in the box,
the foremost one being the lack of space
translational invariance. The detailed treatment of the trapped gas
from the perspective of the present paper, and with particular
attention to the first-order correlation function, will be presented
in a future publication. Here, we just state that the 
the ground state connected correlation function, in place of the constant behaviour of Eq.~(\ref{conn.3})
obeys the Gaussian decay form
\beq
G_{{\rm c},{\bf 0}}({\bf r},-{\bf r})\sim \frac{\exp \{-(r/L)^2 \}}{|{\bf r} - {\bf r}|^{a_0}},
\eeq
where the sector ${\bf r'}=-{\bf r}$ has been considered~\cite{CSSZ} and, again $a_0 = 0$.
The characteristic length $L =\sqrt{\hbar/m_{\textrm{eff}}\Omega}$ represents the linear size of the trap
and the above result, which is in the form of finite-size scaling, indicates that the correlation length is
of order $L$. Thus, in the thermodynamic limit - in the sense of an infinite trap $(L \to \infty)$ - 
a critical state is reached, which is characterised by a non-decaying correlation function, as in the uniform case.
The experimental evidence for the above behaviour is in the results for the first-order correlation function
reported in \cite{Damm}. We stress that this kind of behaviour could not arise in the quasi-average framework,
where the connected correlation function vanishes.

In conclusion, in this paper we have overturned the outlook on the ${\cal GCC}$, from culprit 
responsible of the failure of the GCE to most interesting manifestation of an underlying SSB
pattern characterised by strong amplitude fluctuations and strong correlations.
The proposed conceptual framework enhances the significance of the photon gas system
as the experimental platform where to observe the distinctive features which make
BEC based on the condensation of fluctuations different from BEC based on pure ordering,
as realised in the cold atoms systems.
It seems fitting to end up by saying that the remarkable accomplishment of BEC in GCE conditions 
lends support to the so-called
{\it totalitarian principle}, according to which "everything not forbidden is
compulsory" \cite{Murray,Kragh}.

{\bf Acknowledgements.}
Interesting conversations with Prof. Luca Salasnich, who participated to the early stage of this 
work, are gratefully acknowledged. AS acknowledges
funding from the Italian Ministero dell'Universit\`a e
della Ricerca under the programme PRIN 2022 (''reranking of the final lists''), number 2022KWTEB7, cup
B53C24006470006.

\section{Appendix}

Using the expression~(\ref{GS.4}) for $\rho^{(\nu)}_{\bf 0}$, keeping in mind that
the excited states' contribution $\rho^\prime$ is not affected by the external field and using
the small-$\alpha$ expansion of the Bose function, whereby 
$\lambda^{-3} g_{3/2}(\alpha) =  \lambda^{-3} g_{3/2}(0) - C\sqrt{\alpha} + ...$ 
with $C= 2 \lambda^{-3} \sqrt{\pi}$, 
Eq.~(\ref{BEC.001}) takes the form
\beq
\Delta \rho = \left [\left (\frac{\beta |\nu|}{\alpha} \right )^2 + \frac{1}{V\alpha} \right ] - C\sqrt{\alpha}.
\label{eq.1}
\eeq
It is evident from the two terms in the square bracket that the order of the $V \to \infty$ and $\nu \to 0$ limits matters, 
revealing the singular nature of the perturbation.
The solution, to leading order in $\nu$ and $1/V$, in the two cases is given by
\beq
\alpha^{({\rm qa})} = \left \{ \begin{array}{ll}
(\Delta \rho/C)^2, \quad $for$ \quad \Delta \rho < 0, \\
\\
(\beta|\nu|/\sqrt{C})^{4/5}, \quad $for$ \quad \Delta \rho = 0, \\
\\
\beta|\nu|/\sqrt{\Delta \rho}, \quad $for$ \quad \Delta \rho > 0, 
        \end{array}
        \right . 
\eeq  
  
\beq 
\alpha^{({\rm ra})} = \left \{ \begin{array}{ll}
(\Delta \rho/C)^2, \quad $for$ \quad \Delta \rho < 0, \\
\\
(CV)^{-2/3}, \quad $for$ \quad \Delta \rho = 0, \\
\\
(V\Delta \rho)^{-1}, \quad $for$ \quad \Delta \rho > 0.
        \end{array}
        \right . 
 \label{al.1}
  \eeq
Inserting the above results into Eq.~(\ref{GS.2}) and using Eq.~(\ref{GS.3}), after taking the limits
Eqs.~(\ref{P0.1}) and~(\ref{P0.2}) are obtained.


\begin{thebibliography}{99}



\bibitem{Ziff} R. M. Ziff, G. E. Uhlenbeck and M. Kac, Phys. Rep. {\bf 32}, 169 (1977).

\bibitem{Holthaus} M. Holthaus, E. Kalinowski, and K. Kirsten, Annals of Physics {\bf 270}, 198 (1998).

\bibitem{Fujiwara} I. Fujiwara, D. Ter Haar, and H. Wergeland, J. Stat. Phys. {\bf 2}, 329 (1970).

\bibitem{Grossmann} S. Grossmann and M. Holthaus, Phys. Rev. Lett. {\bf 79}, 3557 (1997).

\bibitem{Weiss} C. Weiss and M. Wilkens, Optics Express {\bf 1}, 272 (1997).

\bibitem{Kruk} M. B. Kruk, P. Kulik, M. F. Andersen, P. Deuar, M. Gajda, K.
Pawłowski, E. Witkowska, J. J. Arlt, and K. Rz aewski,
Rep. Prog. Phys. {\bf 88}, 106401 (2025).

\bibitem{Yukalov} V. Yukalov, Laser Phys. Lett. {\bf 34}, 113001 (2024); V. Yukalov, AVS Quantum Sci. {\bf 7}, 023501 (2025)
and references therein.



\bibitem{Suto} A. S\'uto, Phys. Rev. Lett. {\bf 94}, 080402 (2005).

\bibitem{Lieb} E. H. Lieb, R. Seiringer and J. Yngvason, Phys. Rev. Lett. {\bf 94}, 080401 (2005);
Rep. Math. Phys. {\bf 59}, 389 (2007).

\bibitem{Bogoliubov2} N. N. Bogoliubov, {\it Lectures on Quantum Statistics} vol. 2: {Quasi-Averages} (Gordon and
Breach, New York 1970).

\bibitem{Klaers} J. Klaers, J. Schmitt, F. Vewinger, and M. Weitz, Nature {\bf 468}, 545 (2010).
\bibitem{Schmitt1} J. Schmitt et al., Phys. Rev. Lett. {\bf 112}, 030401 (2014).
\bibitem{Schmitt2} J. Schmitt et al., Phys. Rev. Lett. {\bf 116}, 033604 (2016).



\bibitem{MZ} M. Zannetti, Europhys. Lett. {\bf 111}, 20004 (2015).

\bibitem{CSZ} A. Crisanti, A. Sarracino and M. Zannetti, Phys. Rev. Res. {\bf 1}, 023022 (2019).

\bibitem{Glauber} R. J. Glauber, Phys. Rev. {\bf 131}, 2766 (1963).

\bibitem{Sudarshan} E. C. G. Sudarshan, Phys. Rev. Lett. {\bf 10}, 277 (1963).


\bibitem{Lewis1} H. W. Lewis and G. H. Wannier, Phys. Rev. {\bf 88}, 682 (1952).

\bibitem{Lewis2} H. W. Lewis and G. H. Wannier, Phys. Rev. {\bf 90}, 1131 (1953).

\bibitem{KT} M. Kac and C. J. Thompson , J. Math. Phys. {\bf 18}, 1650 (1977).

\bibitem{YW} C. C. Yan, G. H. Wannier, J. Math. Phys. {\bf 6}, 1833  (1965). 

\bibitem{Castellano} C. Castellano, F. Corberi, and M. Zannetti, Phys. Rev. E {\bf 56}, 4973 (1997).

\bibitem{Fusco} N. Fusco and M. Zannetti, Phys. Rev. E {\bf 66}, 066113 (2002). 

\bibitem{Huang} K. Huang, {\it Statistical Mechanics}, 2nd Edn (Wiley 1987).

\bibitem{BK} T. H. Berlin and M. Kac, Phys. Rev. {\bf 86}, 821 (1952).

\bibitem{Lewis} In this connection it is interesting to note that in the low temperature
phase of the mean-spherical model there are macroscopic fluctuations of the order parameter, which are absent
in the spherical model~\cite{BK} and which are entirely analogous to the ${\cal GCC}$. The interesting aspect is that upon
discovering this Yan and Wannier~\cite{YW} regarded the rersult as  a {\it freak} feature of the model.

\bibitem{Roepstorff}  G. Roepstorff, J. Stat. Phys. {\bf 18}, 191 (1978).

\bibitem{WZ} W. F. Wreszinski and V. A. Zagrebnov, Theor. Math. Phys. {\bf 194}, 157 (2018).

\bibitem{Mehta} C. L. Mehta, Phys. Rev. Lett. {\bf 18}, 752 (1967).

\bibitem{CR} A. Casher and M. Revzen,  Am. J. Phys. {\bf 35}, 1154 (1967).
  

\bibitem{Corberi} F. Corberi, G. Gonnella, A. Piscitelli, and M. Zannetti, J. Phys. A: Math.Theor. {\bf 46}, 042001(2013).
\bibitem{Corberi2}  M. Zannetti, F. Corberi, and G. Gonnella, Phys. Rev. E {\bf 90}, 012143 (2014).
  \bibitem{Corberi3} M. Zannetti, F. Corberi, G. Gonnella, and A. Piscitelli, Commun. Theor. Phys. {\bf 62}, 555 (2014).

\bibitem{Filiasi} M. Filiasi, G. Livan, M. Marsili. M. Peressi, E. Vesselli, and E. Zarinelli, J. Stat. Mech. P09030 (2014).

\bibitem{Ferretti} L. Ferretti, M. Mamino, and G. Bianconi, Phys. Rev. E {\bf 89}, 042810 (2014).

\bibitem{Gunton} J. D. Gunton and M. J. Buckinghan, Phys. Rev. {\bf 166}, 152 (1968).

\bibitem{Reyes} I Reyes-Ayala, F J Poveda-Cuevas, V Romero-Rosh\'in, J. Stat. Mech. 113102 (2019).
  
\bibitem{FCZ1} A. Fierro, A. Coniglio, and M. Zannetti, Phys. Rev. E {\bf 99}, 042122 (2019).

\bibitem{FCZ2} A. Fierro, A. Coniglio, and M. Zannetti, Phys. Rev. E {\bf 102}, 012144 (2020).

\bibitem{CK}
A. Coniglio and W. Klein, J. Phys. A {\bf 13}, 2775 (1980).


\bibitem{CSSZ} A. Crisanti, L. Salasnich, A. Sarracino and M. Zannetti, Entropy {\bf 26}, 367 (2024).

\bibitem{Damm} T. Damm, D. Dung, F. Vewinger, M. Weitz, and J. Schmitt, Nature comm. {\bf 8}, 158 (2017).


\bibitem{Murray} Although there is no clear reference, the quotation is widely attributed to Murray Gell-Mann, see \cite{Kragh}.

\bibitem{Kragh} H. Kragh, arXiv:1907.04623. 


\end{thebibliography}
\end{document}